\documentclass[letterpaper]{jpconf}
\usepackage{graphicx}
\begin{document}
\title{Is Sr$_2$RuO$_4$ a Chiral P-Wave Superconductor?}

\author{C.\ Kallin and A.~J.\ Berlinsky}

\address{Department of Physics \& Astronomy, McMaster University, Hamilton, ON, L8S 4M1, Canada}

\ead{kallin@mcmaster.ca}

\begin{abstract}
Much excitement surrounds the possibility that strontium ruthenate exhibits chiral p-wave superconducting order. Such order would be a solid state analogue of the A phase of He-3, with the potential for exotic physics relevant to quantum computing. We take a critical look at the evidence for such time-reversal symmetry breaking order. The possible superconducting order parameter symmetries and the evidence for and against chiral p-wave order are reviewed, with an emphasis on the most recent theoretical predictions and experimental observations. In particular, attempts to reconcile experimental observations and theoretical predictions for the spontaneous supercurrents expected at sample edges and domain walls of a chiral p-wave superconductor and for the polar Kerr effect, a key signature of broken time-reversal symmetry, are discussed.
\end{abstract}

\section{Introduction}

Unconventional superconductors are those in which superconductivity arises from the direct interaction between particles. In the case of electrons, these interactions may be magnetic.  In $^3$He, they are van der Waals interactions with an attractive tail cut off at short distance by a repulsive atomic core.  Such interactions often favor higher (than s-wave) angular momentum pairing. The cuprate family of high temperature superconductors, which exhibit antiferromagnetic spin fluctuations, is known to have d-wave singlet pairing, while superfluid $^3$He has p-wave triplet pairing. Strontium ruthenate, Sr$_2$RuO$_4$,\cite{sro} is also thought to lie in the class of triplet p-wave superconductors induced by ferromagnetic spin fluctuations.  Crystalline anisotropy affects the detailed nature of the possible pairing states in Sr$_2$RuO$_4$, and a variety of experiments suggest the possibility of a p$_x \pm$ip$_y$ chiral order parameter, which is analogous to the ABM phase of superfluid $^3$He.\cite{mackenzie2003}  Thus Sr$_2$RuO$_4$ has the potential to exhibit the rich order parameter and defect structures of superfluid $^3$He, but for a charged superfluid in a crystalline solid material at temperatures 1000 times higher than for $^3$He.  Some of the interest in strontium ruthenate and chiral p-wave superconductivity stems from the possibility of topologically stable, half-quantum vortices with a single Majorana zero mode bound at the core.\cite{dassarma2006}  Such vortices are expected to exhibit non-Abelian statistics and are potentially  useful for quantum computing because of their topological stability and non-trivial winding properties.

In this paper, we will first review the relevant properties of Sr$_2$RuO$_4$ and briefly discuss the measurements which first suggested that these materials were p-wave triplet superconductors with a chiral gap function of the form p$_x \pm$ip$_y$.  We will then consider in more detail the implications of more recent experiments and the degree to which they do or do not provide a consistent picture for the behavior of Sr$_2$RuO$_4$.  We focus particularly on two experiments, (1) the search, using scanning probes, for magnetic fields associated with the edge currents which are predicted to exist in a chiral p-wave superconductor, as discussed below, and (2) the polar Kerr effect, including new theoretical results by Goryo\cite{goryo} on how impurity scattering can enhance this effect.  In our conclusion, we organize the different experiments into a table summarizing evidence for and against the existence of chiral p-wave superconductivity in Sr$_2$RuO$_4$.

\section{Structure and Properties}

Sr$_2$RuO$_4$ has the same layered perovskite structure as the high temperature superconductor, La$_{2-x}$Sr$_x$CuO$_4$, with Ru replacing Cu. This is the K$_2$NiO$_4$ body-centered tetragonal structure with space group $I4/mmm$. The Ru ion is in a 4+ state $4d^4$ configuration with no net spin, and band overlap leads to metallic behavior with a multi-sheet Fermi surface, as measured by Bergemann et al.\cite{bergemann}  Superconductivity is believed to gap the $\gamma$ sheet which is composed of $d_{xy}$ orbitals oriented in the Ru layers, with induced superconductivity with nodes or a small gap on the two other bands.\cite{rice,gamma,mackenzie2003}.

Sr$_2$RuO$_4$ becomes superconducting below about 1.5K, although the actual $T_c$ is very sensitive to disorder, which is a good indication of an unconventional (non-s-wave) state for which scattering around the Fermi surface can average the gap to zero. Early evidence for the triplet nature of the pairing came from NMR measurements of the Knight shift, which measures the local spin susceptibility and decreases rapidly below $T_c$ for an s-wave superconductor.  For a p-wave superconductor, if the static field is aligned in the plane perpendicular to the d-vector, then the susceptibility looks like that of the normal state and does not change in going through $T_c$. Measurements by Ishida et al.\cite{ishida1998} observed this behavior for fields in the plane of the layers and similar behavior was confirmed by neutron scattering.\cite{duffy}  For triplet pairing, p-wave order is the most likely to occur, although f-wave pairing has not been ruled out experimentally.

In addition to the evidence for triplet pairing, early $\mu$SR measurements of Luke and coworkers gave evidence of broken time reversal symmetry in the superconducting state.\cite{luke1998} In this zero field $\mu$SR experiment, relaxation due to spontaneously created fields arises as $T$ is lowered through $T_c$.  More recent experiments show that the appearance of this relaxation follows $T_c$ as $T_c$ is reduced by the addition of impurities,\cite{luke2008} reinforcing the interpretation that the time-reversal symmetry breaking is directly associated with the superconducting state.

With early experiments on Sr$_2$RuO$_4$ pointing to triplet (most likely p-wave) pairing and broken time reversal symmetry, the question naturally arose of which superconducting order parameters would be  consistent with experiment and with the symmetry of strontium ruthenate.  This problem was studied by Sigrist and others, and summarized in table IV in Mackenzie and Maeno.\cite{mackenzie2003}  There are many possible p-wave order parameters, but if one adds the extra condition of broken time reversal symmetry, there is a single unitary order parameter describing a p-wave state with broken time reversal symmetry.  One would expect the non-unitary states to be less likely to be stabilized in zero magnetic field, as they break the symmetry between up and down spins.  The unitary chiral p-wave state has a uniform gap around the Fermi surface and so is energetically favorable because of the large condensation energy.\cite{nodes}

The order parameter for a p-wave superconductor can be expressed in terms of a d-vector as
\begin{equation}
\Delta({\bf k})=i\{{\bf d}({\bf k})\cdot\vec{\sigma}\}\sigma_y
\end{equation}
where the components of $\vec{\sigma}$ are Pauli matrices and the d-vector contains information about the symmetry of the gap and orientation of the spins.  For unitary (${\bf d}\times{\bf d^*}=0$) states, the spin is zero along the direction of ${\bf d}$.  For example, the d-vector corresponding to ${\bf d}=\Delta_0(k_x + k_y)\hat {\bf z}$ describes a real (except for an overall phase) order parameter with nodes along $k_x = -k_y$ and $<S_z>=0$; the d-vector ${\bf d}=\Delta_0(k_x \hat {\bf y}+ k_y\hat {\bf x})$ corresponds to an order parameter which is real but fully gapped, while ${\bf d}=\Delta_0(k_x \pm ik_y)\hat {\bf z}$ is also fully gapped but has a chirality given by the $\pm$ sign. The latter is the only unitary order parameter describing a p-wave state with broken time reversal symmetry.  Since the two chiralities are degenerate, there is the possibility of metastable domain structures.

This chiral p-wave state is analogous to the A phase of $^3$He.\cite{leggett}  The d-vector here is oriented along the z-axis (chosen to be the c-axis of the crystal), which also corresponds to equal spin pairing in the xy (or ab) plane.  The two chiralities, positive and negative, correspond to the +/- signs in above equation and to Cooper pair wave functions carrying angular momentum plus or minus one along the z-axis.  The BCS wave function for this order parameter carries a total angular momentum of $\hbar$ times the number of Cooper pairs, or $\hbar$ times the number of electrons over 2.\cite{stone2004}

In general, any local perturbation of the chiral p-wave order parameter results in supercurrents.  In particular, spontaneous equilibrium supercurrents are predicted to flow at the edges of a finite sample, confined to within a coherence length of the edge.  This supercurrent is directly related to the angular momentum carried by the state, and is large.\cite{stone2004} However, in a charged superconductor, this current must be screened, so that the magnetic field inside the superconductor vanishes.  Therefore, there is an equal and opposite screening current confined approximately to within the penetration depth plus the coherence length of the surface.  Due to the different spatial distributions of these two currents, there is a net magnetic field at the surface, which is predicted to have a maximum value of about 10 Gauss in an idealized model.\cite{matsumoto} Similar supercurrents and fields result at domain walls.  The field alternates in sign across a domain wall and achieves a maximum magnitude of about 20 Gauss (again, in an idealized model).\cite{volovik1985,matsumoto} The $\mu$SR results described above have been interpreted as evidence for fields associated with internal domain walls.\cite{luke1998}.

Some of the current interest in strontium ruthenate and chiral p-wave superconductivity comes from possible connections to quantum computing.  While the d-vector in strontium ruthenate is believed to be pinned to the c-axis by spin-orbit coupling, the exotic physics relevant to quantum computing, would result from a state with the d-vector free to rotate in the ab plane.   To see this in a simple way, consider a vortex circulating in the ab plane.  If the d-vector rotates around this vortex, a phase of $\pi$ is acquired.  Therefore, the orbital part of the Cooper pair wave function only needs to acquire an additional phase of $\pi$ (not 2$\pi$) which implies that the vortex carries one half of the usual superconducting flux quantum.  One can show that such vortices obey non-Abelian statistics, which is exactly what is required in quantum computing, because of the non-trivial windings and topological stability.\cite{ivanov}

More recent NMR experiments caused some excitement, because they suggest the possibility that the d-vector can be rotated  into the ab plane of strontium ruthenate by applying a magnetic field along c.\cite{murakawa}  While earlier NMR measurements were done with a magnetic field in the ab plane, more recent ones were performed with a field oriented along the c-axis.  In this case, for the chiral p$_x \pm$ip$_y$ state, one would expect to see a drop in the Knight shift as the temperature is lowered below $T_c$ as the spins condense into a state with $\langle S_c\rangle =0$.  Although the NMR linewidth is broader in this field orientation, within the error bars no suppression of the Knight shift is seen in the superconducting state.  This has been interpreted as evidence that the d-vector is rotated into the ab plane by fairly modest fields along c.  However, even if the d-vector is, in fact, rotated to lie perpendicular to the field, the system is not guaranteed to stay in a chiral p-wave state, since there is a non-chiral p-wave state, with the d-vector in the ab-plane, which is energetically competitive with chiral p-wave and which may be stabilized by a field.\cite{annett}

\section{More recent evidence for chiral p-wave}

We now turn to the more recent experimental results on strontium ruthenate.  First, we briefly discuss the phase sensitive measurements, as this method provides detailed information about the order parameter symmetry.  Ying Liu's group used a SQUID geometry which was sensitive to the parity of the wave function, and found compelling evidence for odd pairing, which is compatible with p-wave pairing.  They also found evidence for time reversal symmetry breaking in one corner junction measurement.\cite{nelson2004}  More recently, Dale van Harlingen's group performed tunneling measurements in a Josephson junction geometry.\cite{kidwingara2006}  They observe complicated Fraunhoffer-like patterns, which they could not fit with the two-domain structure of chiral p-wave.  However, by introducing four domains ($p_x \pm ip_y$ and $p_y \pm ip_x$), and allowing for small (of order one micron) dynamic domains, they found better fits to their data.  Since this modeling brings in relative phases of $\pm\pi/2$, it is sensitive to the chirality, or time reversal symmetry breaking.  While only two types of domains are stable in chiral p-wave,  it may be possible to have nontrivial phases, as needed to model the Josephson data, if there are domain walls intersecting the surface at angles other than 90 degrees.\cite{sigristunpublished}  Consequently, these measurements provide intriguing evidence pointing toward time reversal symmetry breaking.

The polar Kerr effect is a direct probe of chirality.  In this measurement, linearly polarized light, incident on the surface of the sample, is reflected as elliptically polarized light with an angle of rotation referred to as the Kerr angle.  In a ferromagnet, the Kerr angle is directly related to the magnetization perpendicular to the surface.  Kapitulnik and coworkers  observed a low temperature Kerr angle, with a probing frequency of 0.8eV, of approximately 65 nanoradians in the low temperature superconducting state of strontium ruthenate.\cite{xia2006} Cooling in a magnetic field affected the sign of the Kerr angle, but not the magnitude, which would be consistent with observing the response of a single domain, whose chirality was determined by the direction of the field.

The Kerr angle is related to the Hall conductivity and is proportional to the imaginary part of the Hall conductivity in the limit of large frequency.  In this limit, the Kerr effect can be understood as the system preferentially absorbing light with positive (or negative) circular polarization, depending on the chirality.  The Hall conductivity of a chiral p-wave state has been extensively studied in the clean limit.\cite{kerrtheory}  It follows from translational symmetry, that an ideal, clean chiral p-wave superconductor has vanishing Hall conductivity and Kerr angle.\cite{read}  Interesting effects do occur at finite wavevector, where one finds a substantial  Hall conductivity.  It would be interesting (although difficult!) to probe strontium ruthenate at finite wave vector to see if this large signature of chirality, even in a clean system, is observed.

Recently, Goryo\cite{goryo} identified the leading order contribution to the polar Kerr effect due to disorder.  The usual lowest order terms, proportional to $n_iU^2$, vanish, but the ``skew-scattering" diagrams, proportional to $n_iU^3$, where a given impurity scatters 3 times, give a contribution which Goryo estimates to be between 5 and 30 nanorads under certain assumptions, not too much smaller than the observed maximum value of 65 nanorads.  A more realistic treatment of the impurity potential and mixing in of other harmonics would decrease the calculated impurity induced contribution.  In addition, the probing frequency of 0.8eV is so much larger than the gap, $\sim$0.1 meV, that one might not be able to ignore the energy dependence (cutoff) of the pairing potential.  Nevertheless, it is intriguing that this impurity contribution gives a result not too different from experiment, and it merits further investigation.  The calculation predicts an unusual $\omega^{-4}$ dependence (from the extra factor of U) which could be checked experimentally.  Furthermore, the experimental result should depend strongly on the amount of disorder within the optical skin depth of the ab surface.

Finally, we turn to scanning SQUID microscopy which can probe the fields localized at the edges of the ab surface.\cite{kwon}  Fields at domain wall boundaries which intersect the ab surface should also be detectable by this technique.  These measurements were carried out by Kirtley,\cite{kirtley} who found no strong signals other than those from trapped vortices.  Fig.1 illustrates this fact, where the data from a SQUID scan across the ab surface is shown as the solid line near zero flux.  The dotted line which lies very close to the data is the result of modeling a conventional s-wave superconductor in the presence of an external field of 3 nT.  The small feature at the surface can be accounted for by this small residual field.  By contrast, the signal one would expect from modeling ideal surface currents, as described in Matsumoto and Sigrist,\cite{matsumoto} modified for field fringing effects at the surface, is shown by the dashed line which extends well off this linear scale.  The expected edge currents would have to be approximately two orders of magnitude smaller to explain this null observation.
Similar experiments using a scanning Hall probe also saw no surface fields except those due to trapped vortices.\cite{kirtley,bjornsson2005}

One can model the effect of domain walls intersecting the ab surface which would give rise to additional surface fields as one scans across the ab face.  However, since the field alternates in sign across a wall, if the domain size becomes too small, the signal is reduced because of the finite size of the Hall probe or SQUID pickup loop.  Domain sizes of 1-2 microns extending to the edges of the ab face could explain the null results, even if the supercurrents were of the predicted size for an ideal chiral p-wave superconductor.

\begin{figure}
\begin{center}
\includegraphics[width=4.5in]{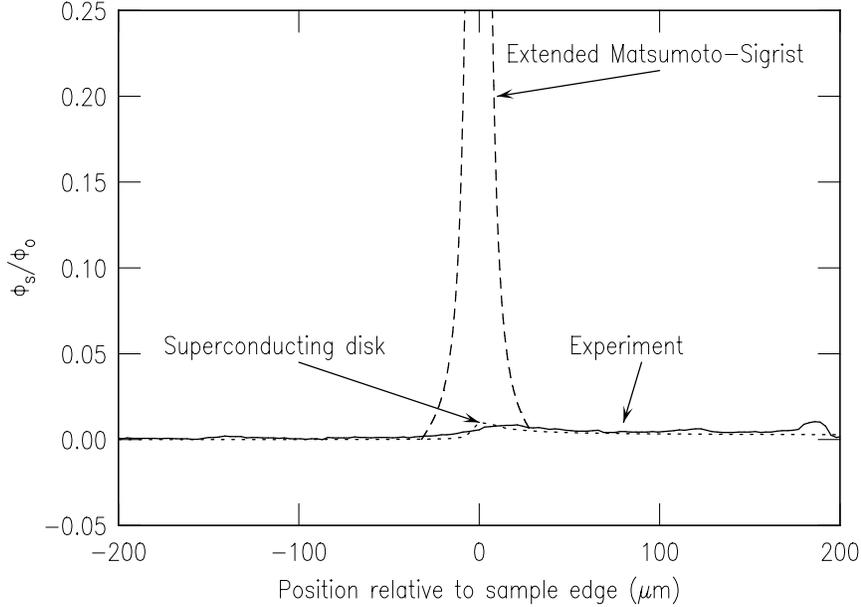}
\vspace{0.1in}
\caption{
SQUID scan across the edge of an ab face of a  Sr$_2$RuO$_4$ crystal (solid line). The dotted line is the prediction for an s-wave superconducting disk
in a uniform residual field of 3 nT. The dashed line is the prediction for a single domain
$p_x+ip_y$ superconductor, following the theory Matsumoto and Sigrist, but modifed for a finite sample. The peak value of the dashed line is 1.  (From Kirtley et al., Ref.\cite{kirtley}.)
}
\label{fig:srocrs}
\end{center}
\end{figure}

One can think of many effects which could reduce the magnitude of the edge supercurrents, although most of these would reduce the magnitude of all spontaneous supercurrents and, therefore, would be difficult to reconcile with the muon spin resonance experiments which are interpreted as evidence for fields associated with internal domain walls.  For example, multiband effects, anisotropy, and disorder can reduce the magnitude of all spontaneous supercurrents.  Similarly, Leggett has proposed an alternative to the BCS chiral p-wave wave function, which he argues would reduce the spontaneous edge currents by a factor of $(\Delta/E_f)^2$.\cite{leggettwf}  This would certainly put the edge fields below the limit of observability for scanning SQUID and Hall probes, but it would also put it below that of $\mu$SR.  Similarly, other pairing wave functions, as might be captured by chiral p-wave Ginzburg-Landau theory with parameters quite distinct from those derived from weak-coupling for a simple pairing Hamiltonian, would affect all supercurrents as well as the $\mu$SR measurements.

One effect that is specific to edges is surface roughness and other surface pairbreaking effects.  However, Ginzburg-Landau calculations of such surface effects show that they only reduce the current by at most a factor of two, and not by the orders of magnitude needed to explain the null scanning results.\cite{ashby}  Nucleating other order parameters at the surface can have a greater effect on the surface currents, and even lead to a magnetic field at the surface which changes sign.\cite{ashby}  This could be compatible with both a postive $\mu$SR signal and a null scanning SQUID or Hall probe result, but would imply that the tunneling measurements,\cite{nelson2004,kidwingara2006} which have been taken as evidence for chiral p-wave, were not measuring the bulk order parameter.  It is difficult to imagine effects which would only affect the scanning SQUID and Hall probe measurements, while leaving the various positive experimental results from other techniques intact.

\section{Conclusions}
\begin{table}[h]
\caption{\label{trsb}Time Reversal Symmetry Breaking}
\begin{center}
\begin{tabular}{lll}
\br
Experiment&TRSB?&Estimated Domain Size \\
\mr
muSR &Yes&$< 2\mu$\\
Polar Kerr Effect&Yes&$> 50\mu$ ($15-20\mu$) with (without) field cooling\\
Scanning Hall Probe &No&$< 1\mu$\\
Scanning SQUID &No&$< 2\mu$\\
Josephson Junction Tunneling &Yes&$< 1\mu$ ($~0.5\mu$ dynamic)\\
SQUID Tunneling &Parity&$> 10-50\mu$\\
SQUID Corner Junction&Yes&$ >10\mu$ \\
\br
\end{tabular}
\end{center}
\end{table}

Table I summarizes the current experimental situation with respect to time reversal symmetry breaking in the superconducting state of strontium ruthenate.  While there are a variety of different probes giving positive results, the situation is less clear when one looks more closely at the details.  For example, note that all these measurements either invoke domain walls to explain their positive results (Josephson junction tunneling and $\mu$SR) or require few or no domain walls (SQUID tunneling, polar Kerr effect).  It is important to note that domain walls cost energy and under ideal circumstances the ground state would be a single domain.  However, in the presence of disorder and defects, domains which are nucleated as the sample is cooled through the transition, may be pinned.  Cooling in a field should reduce domains by biasing the system to one chirality, but would then leave pinned vortices which can also affect many measurements.  Currently, taken at face value, the various experiments appear incompatible with each other in the assumptions they require about domain walls density, as seen from Table I.  While different sample treatment may explain some of this discrepancy, clearly one would like to see more systematic studies and attempts to control domain wall densities.  Ideally, one would like to directly observe domain walls if, in fact, strontium ruthenate is a chiral p-wave superconductor.

In summary,  our view is that strontium ruthenate is an unconventional superconductor and most likely exhibits triplet pairing, although further work is needed in accurately describing the Knight shift data for different field directions.  There is substantial and intriguing evidence for spontaneous time reversal symmetry breaking, but puzzles remain in trying to connect this with chiral p-wave order.  In addition to the apparent discrepancies of domain wall densities, there is also the details of muSR which does not see any strong signal of the muons' own induced field, as expected for chiral p-wave.  Similarly, the relatively large size of the observed Kerr angle is surprising, although the recent calculations of Goryo suggest further work to be done in determining whether substantial disorder explains the magnitude.  For the discrepancy in domain size, it would be of great interest to try to control and directly detect domain walls in this state.  Finally, as in $^3$He-A, direct observation of the macroscopic angular momentum remains elusive.  It would be of interest to develop new probes to look for the corresponding edge supercurrents, such as low temperature beta-NMR and slow neutrons.  Scanning tunneling microscopy would also be of great interest, to see if the bulk ab face differs from the ab edges, although surface reconstruction is a problem for this technique.  If the edge supercurrents remain elusive and the other evidence of chiral p-wave symmetry is firmed up,  one may need to further investigate models which might describe a chiral p-wave superconductor without the large angular momentum.  Such theories are likely to be relevant to $^3$He as well.

\section{Acknowledgements}

We would like to acknowledge the hospitality of the Stanford Institute for Theoretical Physics where this study began and the Kavli Institute for Theoretical Physics for their hospitality and, in particular, for their support of the December 2007 ``Sr$_2$RuO$_4$ and Chiral p-wave Superconductivity Miniprogram" where many of the issues presented in this paper were thoroughly discussed. We would also like to thank  Kathryn Moler, John Kirtley, Clifford Hicks, Eun-Ah Kim, Yoshi Maeno,  Manfred Sigrist, Victor Yakovenko, Roman Lutchyn, Aharon Kapitulnik, Graeme Luke, Rahul Roy, Sung-Sik Lee and Philip Ashby for many useful discussions.  We are grateful to the Natural Sciences and Engineering Research Council of Canada and to the Canadian Institute for Advanced Research for their continuing support.


\begin{thebibliography}{9}
\bibitem{nodes} The experimental evidence for and against nodes in the gap is ambiguous.  The specific heat, $1/T_1$, the thermal conductivity and the temperature dependence of the penetration depth, suggest that there are nodes in the gap, which would not be the case for simple p$_x \pm$ip$_y$ which is fully gapped.  However, experiments at lower temperatures are needed to distinguish between possible nodes and small gaps on parts of the multi-sheet Fermi surface.
\end{thebibliography}
\end{document}